\documentclass[pra, aps, 10pt, twocolumn,floatfix]{revtex4-1}
\usepackage{amsmath,graphicx}
\usepackage[pdftex,plainpages=false,colorlinks=true,linkcolor=blue, citecolor=blue, urlcolor=blue]{hyperref}
\usepackage[charter]{mathdesign}
\usepackage{bm}
\usepackage[x11names,svgnames]{xcolor}

\newcommand\rv{\mathbf{r}}

\newcommand\Qv{\mathbf{Q}}

\newcommand\sv{\mathbf{s}}
\newcommand\Sv{\mathbf{S}}
\newcommand\Sigmav{\bm{\Sigma}}
\newcommand\Tau{\bm{\tau}}

\newcommand\Gv{\mathbf{G}}
\newcommand\Tr{\mathrm{Tr}}

\begin{document}
\title{ Phase diagram of the Hubbard-Kondo lattice model from variational cluster approximation}
\author{J. P. L. Faye}
\affiliation{The Abdus Salam International Center for Theoretical Physics, Strada Costiera 11, 34014 Trieste, Italy}
\author{ P. Ram}
\author{B. Kumar}
\affiliation{School of Physical Sciences, Jawaharlal Nehru University, New Delhi 110067, India}

\author{D. S\'en\'echal}
\affiliation{D\'epartement de physique and Institut Quantique, Universit\'e de Sherbrooke, Sherbrooke, Qu\'ebec, Canada J1K 2R1}
\author{M. Kiselev}
\affiliation{The Abdus Salam International Center for Theoretical Physics, Strada Costiera 11, 34014 Trieste, Italy}
\date{\today}
\begin{abstract}
The interplay  between the Kondo effect  and magnetic ordering driven by  the Ruderman-Kittel-Kasuya-Yosida interaction is studied within the 
two-dimensional Hubbard-Kondo lattice model. In addition to the antiferromagnetic 
exchange interaction, $J_\perp$, between the localized and the conduction electrons,  this model also contains the local repulsion, $U$, between the conduction electrons. We use  
variational cluster approximation  to investigate the competition 
between the antiferromagnetic phase, 
 the Kondo singlet phase, and  a ferrimagnetic phase on square lattice.
At half-filling, the N\'eel antiferromagnetic phase 
dominates from small to moderate $J_\perp$ and $UJ_\perp$, and the Kondo singlet elsewhere.  Sufficiently away from half-filling, the antiferromagnetic  phase 
first gives way to a ferrimagnetic phase (in which the localized spins order ferromagnetically, and the conduction electrons do likewise, but the two mutually align antiferromagnetically), and then to the Kondo singlet phase.
\end{abstract}
\maketitle
\section{Introduction}\label{Sec:Introduction}
The interaction between itinerant electrons and impurity spins plays a key role in many areas of condensed matter physics, including, but not limited to, quantum materials, spintronics, and quantum information processing~\cite{Childress281}.
In quantum materials, this interaction can arise either through (i) the hybridization of valence electrons with localized $d$ or $f$ orbitals or (ii) a coupling of the electron spin density to the spins of the localized electrons.
In the first case, it can be argued that the Kondo exchange~\cite{Kondo:1964} becomes the dominant interaction if the localized orbitals, with a weak hybridization, are slightly occupied~\cite{PhysRev.149.491}. In the second case, corresponding to half-filled local orbitals, the Kondo lattice model 
 presents a generic  description of the low-energy physics.
These two mechanisms describe the physics of 
two main families of the strongly correlated heavy fermion (HF) 
systems: In uranium-based HF systems, the $5f$ electrons are strongly hybridized with $s$, $p$ or $d$ itinerant electrons. 
As a result, there exist strong charge (valence) fluctuations. 
By contrast, the $4f$ level in Cerium-based HF systems is located well below the Fermi level, due to which
the charge fluctuations are frozen out and the spin fluctuations play the central role. 
The effective model describing their low-energy physics is known as the Kondo lattice model~\cite{Kondo:1964,Prasanta:2008, coleman:2015, hasan:2015}.

The ground state of  the Kondo lattice model at half-filling is insulating either due to the formation of singlets between the local moments and the cloud of conduction electrons (Kondo cloud)~\cite{PhysRevB.76.115108}, or due to magnetic ordering of local moments via the Ruderman-Kittel-Kasuya-Yosida (RKKY) mediated by itinerant electrons~\cite{PhysRevB.51.15630, PhysRevLett.83.796, PhysRevB.64.092406, PhysRevB.56.11820, coleman:2015}.  The 
mean field theory reveals that the magnetic correlations  depend on the density of conduction electrons: they are  antiferromagnetic (AFM) near half-filling and ferromagnetic (FM) at lower fillings~\cite{PhysRevB.20.1969}. 
Increasing the exchange interaction 
between the conduction electrons and the localized moments leads the magnetic system to the spin-gapped Kondo singlet phase. The  spin gap formation associated with hidden symmetries has been investigated analytically in  spin chains with AFM Heisenberg exchange interactions coupling the conduction and the localized spins~\cite{kiselev:2005}. This study was supported by Monte Carlo simulations~\cite{PhysRevLett.100.017202, PhysRevB.82.174410}. Notice that these chains called spin-rotor chains are similar to the spiral staircase Heisenberg Ladder~\cite{Doniach1987} for the study of Kondo physics.

The Kondo lattice model of noninteracting conduction electrons is the most promising candidate to capture  
the qualitative physics of the HF systems, but it fails to correctly describe  the physics at lower temperature scale~\cite{Fulde1993}. One example is the electron-doped cuprate $\mathrm{Nd}_{2-x}\mathrm{Ce}_x\mathrm{CuO}_4$~\cite{PhysRevLett.71.2481},  wherein it is suggested that the Kondo  effect plays some important role due to strong correlation among  the charge carriers and therefore cannot be neglected. Indeed, the  interaction between conduction electrons can play a crucial role and even enhance the Kondo temperature significantly~\cite{Fulde1993}. 
In this spirit, some effort has been undertaken  to study the influence of the correlated conduction electrons on the Kondo effect both for impurity~\cite{PhysRevLett.72.892, PhysRevB.52.R6979, PhysRevB.53.3211, PhysRevB.53.5626} and lattice models~\cite{PhysRevB.53.R8828, PhysRevB.54.R752, PhysRevB.56.6559}. 
In the Kondo lattice model, the interaction is introduced by adding to the Hamiltonian a Hubbard type repulsion, $U$, among the conduction electrons. 
A resulting Anderson-Hubbard model was shown to map into an impurity model~\cite{PhysRevB.56.6559} within dynamical mean field theory~\cite{RevModPhys.68.13, Pruschke}; the impurity consisted of two correlated orbitals. 
In Ref.~\cite{PhysRevB.59.9888}, the role of this Coulomb repulsion was investigated using both $T = 0$ Quantum Monte Carlo and a bond-operator mean field theory at half-filling. One of their findings 
is that this model displays a  magnetic order-disorder transition  
with a critical Kondo interaction which decreases as a function of the Hubbard repulsion.

In this paper, we study the interplay between the Kondo effect and the magnetic  ordering within the Hubbard-Kondo lattice model.
It includes a local Coulomb repulsion, $U$, between the conduction electrons, in addition to the AFM Kondo interaction, $J_\perp$. 
We obtain  its quantum phase diagram at half-filling, and also at finite dopings, using the Variational Cluster Approximation (VCA)~\cite{Dahnken:2004, Potthoff:2012fk,Potthoff:2014rt}. The VCA is an approach based on rigorous variational principle that treats short-range correlations exactly. At half-filling, we find that the ground state is a N\'eel AFM at moderate to small values of $J_\perp$ and $UJ_\perp$,  while the Kondo singlet phase is stable at large $J_\perp$ and $U$.  The transition from the AFM to the Kondo singlet phase is continuous (second-order). 
At finite doping, we find that the antiferromagnet survives close to half-filling and disappears upon doping further or increasing $U$ (at least for small $U$). A ferromagnetic phase becomes stable at lower density and small exchange interactions. 
The Kondo singlet appears at large   $J_\perp$ for all values of the   conduction electron density. 
The transition from the  magnetically ordered to the Kondo singlet phase becomes discontinuous (first order) away from half-filling.

The paper is organized as follows. In Sec.~\ref{Sec:Model}, we define the model and briefly review the VCA method. We present and discuss our results in Sec.~\ref{Sec:Results}, and conclude in Sec.~\ref{conclusion}.
\begin{figure}
\centerline{\includegraphics[scale=0.6]{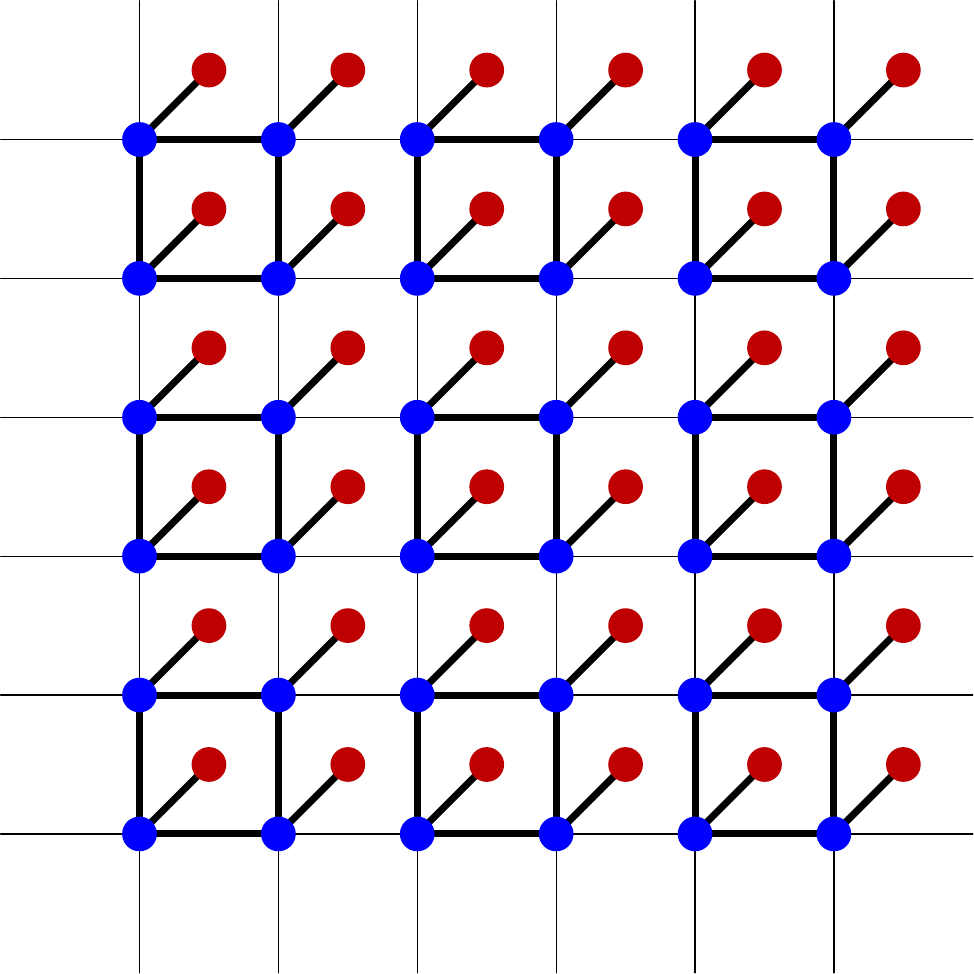}}
\caption{The Hubbard-Kondo lattice and its decomposition  into identical 8-site clusters.
The conduction and localized orbitals are represented by blue and red dots, respectively.\label{fig:lattice}}
\end{figure}
\section{Model and method} \label{Sec:Model}
The Kondo lattice model (or the necklace Hamiltonian) was introduced by Doniach~\cite{Doniach1987}.
It can be formulated as a tight-binding model of the conduction electrons on a lattice,  where on each site there also sits a local spin that couples to the conduction electron spin.
The Kondo-Hubbard model is obtained by adding a local Coulomb repulsion $U$ to the conduction electrons.
In addition, the local spins may be represented by a half-filled band of localized electrons (the $f$-band)
with a very strong local repulsion $U_f$.
\subsection{Model Hamiltonian}
The Hubbard-Kondo Hamiltonian on square lattice can be written as follows: 
\begin{eqnarray} \label{H_Kondo}
H =  H_{\text{K}} +  H_{\text{H}} 
\end{eqnarray}
where the Kondo part is
\begin{eqnarray} 
\nonumber H_{\text{K}} &=& - t\sum_{\langle i,j \rangle\sigma}c^{\dagger}_{i \sigma}c_{j \sigma} - t^{\prime}\sum_{\langle\langle i,j \rangle\rangle\sigma}c^{\dagger}_{i \sigma}c_{j \sigma}\\ && \nonumber  -\mu\sum_{i\sigma}n_{i\sigma}  -\mu_f\sum_{i\sigma}n^f_{i\sigma} + J_{\perp} \sum_{\langle i,j \rangle}  \sv_i   \cdot \Sv_j 
\end{eqnarray}
and the Hubbard part is:
\begin{eqnarray} 
\nonumber H_{\text{H}} = U\sum_{i}n_{i\uparrow}n_{i \downarrow} +  U_f\sum_{i}n^f_{i\uparrow}n^f_{i \downarrow} 
\end{eqnarray}
In the above,  $c_{i \sigma}$ annihilates a conduction electron at site $i$ with spin $\sigma$, $t$ is the nearest-neighbor and $t^{\prime}$  the next-nearest-neighbor hopping amplitude; $\mu$  and $\mu_f$ are the chemical potentials for the conduction and the localized electrons, respectively; likewise, $U$ and $U_f$ are their onsite repulsions.
The  number of  conduction electrons at site $i$ with spin $\sigma$ is $n_{i\sigma} = c^{\dagger}_{i \sigma}c_{i \sigma}$, and likewise $n^f_{i\sigma} = f^{\dagger}_{i \sigma}f_{i \sigma}$ for localized electrons. 
The $J_{\perp}$ is the exchange interaction between the itinerant spins, $\sv_i =  \frac{1}{2}c^{\dagger}_{i \sigma}\Tau_{\sigma\sigma^{\prime}}  c_{i \sigma^{\prime}}$, and the localized spins $\Sv_j = \frac{1}{2}f^{\dagger}_{i \sigma}\Tau_{\sigma\sigma^{\prime}}  f_{i \sigma^{\prime}}$, where $\Tau_{\sigma\sigma^{\prime}}$ are the Pauli matrices. 
We assume an AFM coupling: $J_{\perp} > 0$. In this paper, we will define all quantities in units of the hopping amplitude $t$, that is, set $t = 1$.

The $f$-electrons are truly local here due the absence of hopping among the $f$-electrons and the hybridization between the conduction and the $f$-electrons. We take $\mu_f = U_f/2$, and set $U_f$ to a value ($100$) much larger than any other parameter of the model. This makes sure that we have exactly one $f$-electron (i.e., a localized spin-1/2 moment) per site. In the limit $J_{\perp}\gg t$, the ground state can be shown to be a product of the singlets formed locally between the conduction electrons and the localized moments. 
For the conduction electrons at half-filling, their chemical potential is $\mu = U/2$. By varying $\mu$ away from this value, we achieve different dopings for the conduction electrons.

\subsection{The Variational Cluster Approximation}\label{Sec:VCA}
In order to probe the possibility of magnetism in model~\eqref{H_Kondo}, we use the variational cluster approximation (VCA) with an exact diagonalization solver at zero temperature~\cite{Dahnken:2004}.
This method has been applied to many strongly correlated systems in connection with various broken symmetry phases, for example  in superconductivity~\cite{Senechal:2005,Aichhorn:2006} and magnetism.\cite{Nevidomskyy:2008}
For a detailed review of the method, see Refs~\cite{Potthoff:2012fk,Potthoff:2014rt}.
Like other quantum cluster methods, VCA starts by a tiling of the lattice into an infinite number of  identical clusters. 
We will use the 8-site cluster illustrated in Fig.~\ref{fig:lattice}.
In VCA, one considers two systems: the original system described by the Hamiltonian $H$, defined on the infinite lattice, and the {\it reference system}, governed by the Hamiltonian $H'$, defined on the cluster only, with the same interaction part as $H$.
Typically, $H'$ will be a restriction of $H$ to the cluster (i.e., with inter-cluster hopping removed), to which various Weiss fields may be added in order to probe broken symmetries.
More generally, any one-body term can be added to $H'$.
The size of the cluster should be small enough for the electron Green function to be computed numerically, by an exact diagonalization method.
The optimal one-body part of $H'$ is determined by a variational principle. More precisely, the electron self-energy $\Sigmav$ associated with $H'$ is used as a variational self-energy, in order to construct the Potthoff self-energy functional~\cite{Potthoff:2003b}:
\begin{multline}\label{eq:omega}
\Omega[\Sigmav(\xi)]=\Omega'[\Sigmav(\xi)]\\ +\Tr\ln[-(\Gv^{-1}_0 -\Sigmav(\xi))^{-1}]-\Tr\ln(-\Gv'(\xi))
\end{multline} 
The quantities $\Gv'$ and $\Gv_0$ above are the physical Green function of the cluster and the non-interacting Green function of the lattice, respectively. 
The symbol $\xi$ stands for a small collection of parameters that define the one-body part of $H'$. 
$\Tr$ is a functional trace, i.e., a sum over frequencies, momenta and bands, and $\Omega'$ is the grand potential of the cluster, i.e., its ground state energy, since the chemical potential $\mu$ is included in the Hamiltonian.
$\Gv'(\omega)$ and $\Omega'$ are computed numerically via the Lanczos method at zero temperature.
\begin{figure}
\centerline{\includegraphics[width=\hsize]{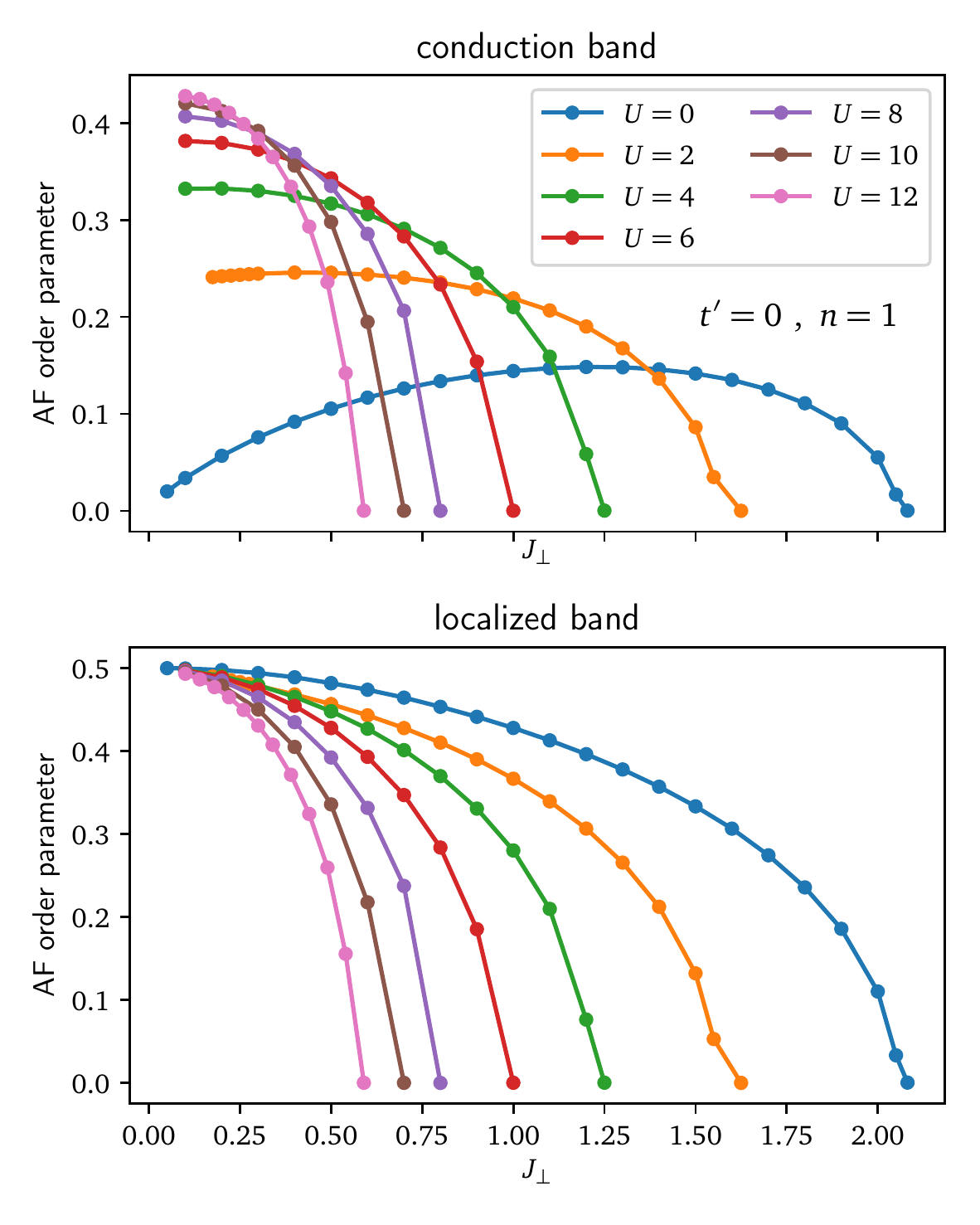}}
\caption{{  The} AFM order parameter of the conduction electrons (upper panel)  and localized $f$-electrons (lower panel) as a function of $J_{\perp}$ at half-filling ($n = 1$) for several values of the on-site repulsion $U$ ranging from 0 to 10; from the Variational Cluster Approximation. See text for details.\label{fig:AFM}}
\end{figure}
    
The Potthoff functional $\Omega[\Sigmav(\xi)]$ in Eq.~\eqref{eq:omega} is computed exactly, but on a restricted space of the self-energies $\Sigmav(\xi)$ that are the physical self-energies of the reference Hamiltonian $H'$.
We use a standard optimization method (e.g. Newton-Raphson) in the space of parameters $\xi$ to find the stationary value of $\Omega(\xi)$:
\begin{equation}\label{eq:Euler}
\frac{\partial\Omega(\xi)}{\partial\xi} = 0
\end{equation}
This represents the best possible value of the self-energy $\Sigmav$, which is used, together with the non-interacting Green function $\Gv_0$, to construct an approximate Green function $\Gv$ for the original lattice Hamiltonian $H$.
From that Green function one can compute the average of any one-body operator, in particular {  the} order parameters associated with magnetism.
The actual value of $\Omega(\xi)$ at the stationary point is a good approximation to the physical grand potential of the lattice Hamiltonian $H$.
\begin{figure}
    \centerline{\includegraphics[width=0.9\hsize]{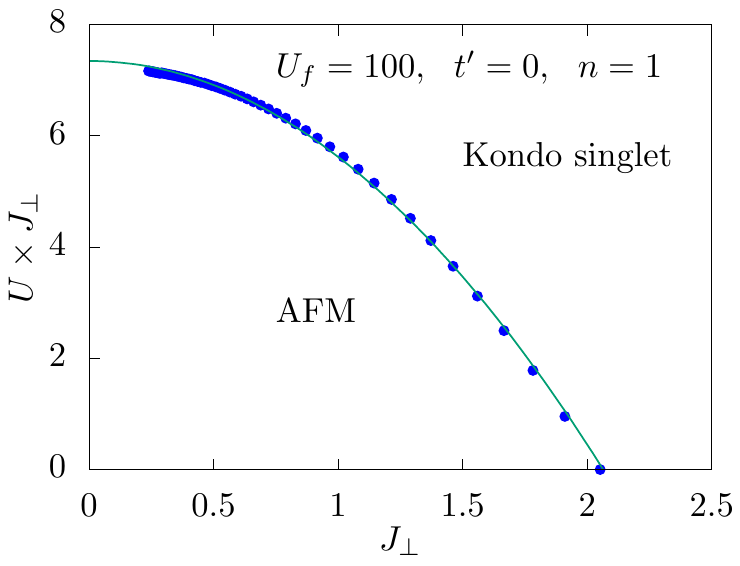}}
    \caption{Phase diagram of the Kondo-Hubbard model~\eqref{H_Kondo} at half-filling ($n = 1$)  and $t^{\prime} = 0$ in the $J_{\perp}$-$UJ_{\perp}$ plane. 
    Note how the critical $U$ towards the Kondo singlet scales like $1/J_\perp$ when $J_\perp$ is small. The green curve is the quadratic fit $J_\perp (J_\perp + a U) = b$ given by  the theory of Kondo insulators  where from our data  $a\approx 0.58$ and $b\approx 4.26$.   \label{fig:phase_diagram_HF}}
\end{figure}
    
There may be more than one stationary solutions to Eq.~\eqref{eq:Euler}. For instance: A {\it normal state\/} solution in which all Weiss fields used to describe broken symmetries are zero, and another solution, with a non-zero Weiss field, describing a broken symmetry state. 
As an additional principle, we assert that the solution with the lowest value of the functional~\eqref{eq:omega} is the physical solution~\cite{Potthoff:2006mb}.
Thus competing phases may be compared via their value of the grand potential $\Omega$, obtained by introducing different Weiss fields.
\section{Results and discussion}\label{Sec:Results}
In order to probe magnetism in model~\eqref{H_Kondo}, we introduce the following local operators {  (for the conduction as well as $f$-electrons)} in the cluster Hamiltonian, within the VCA:
\begin{eqnarray} 
\hat M_\Qv = M_\Qv \sum_i e^{i \Qv \cdot \rv_i} (n_{i\uparrow}-n_{i\downarrow})
\end{eqnarray}
where $\Qv = (\pi,\pi)$ for antiferromagnetism and $(0,0)$ for ferromagnetism, {  and} $M_\Qv$ is the Weiss field, which is determined by solving Eq.~\eqref{eq:Euler} ($\xi=M_\Qv$).
We have applied the VCA with the cluster system shown in Fig.~\ref{fig:lattice}, and {  used} the AFM Weiss field $M_{(\pi,\pi)}$ at half-filling and both $M_{(\pi,\pi)}$ and {  $M_{(0,0)}$} away from half-filling.
\subsection{Phase diagram at half-filling}\label{pdhf}
Figure~\ref{fig:AFM} shows the AFM order parameter as a function of $J_\perp$ for several values of $U$, obtained {  from} the VCA by using a single variational parameter, $M_{(\pi,\pi)}$, for antiferromagnetism in the conduction band.
The second-neighbor hopping $t^{\prime}$ is set  to zero in order to  fulfill   particle-hole symmetry. 
The upper panel shows the AFM order parameter in the conduction band and the lower panel the corresponding quantity in the localized band.
Upon increasing $J_\perp$, the system undergoes, as expected, a continuous transition from an AFM phase to a Kondo singlet phase at some critical value of $J_\perp$.
This critical value decreases upon increasing $U$, and therefore the singlet phase is favored by the on-site interaction.
\begin{figure}
\centerline{\includegraphics[scale=0.9]{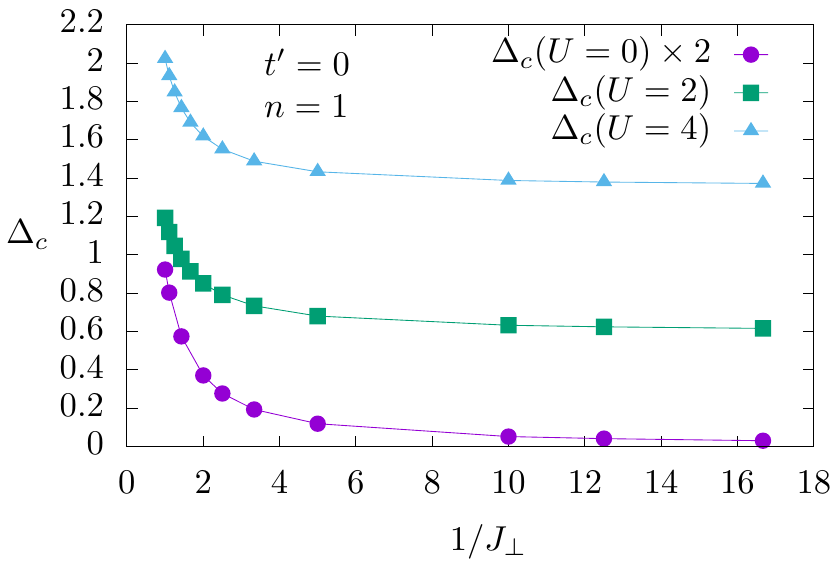}}
\caption{The charge gap as a function of $1/J_\perp$  at half-filling for $U = 0, 2, 4$. 
The second neighbour hopping $t^{\prime}$ is set to 0. 
The system is an insulator for any finite value of $J_{\perp}$.\label{fig:gap}}
\end{figure}
The exchange interaction alone can be responsible for both magnetic and Kondo singlet phases depending on its strength, as we can see from the $U=0$ curve. For $U=0$, the critical exchange interaction is found to be   $J_{\perp} = 2.05$, quite a bit larger than the value $J_{\perp} = 1.45$ found using the Monte Carlo method~\cite{PhysRevB.70.020402}. 
This can be attributed to the small cluster size, which quenches the destabilizing action of spin waves, which can only act within the cluster itself. In comparison, the calculations  in Ref.~\cite{PhysRevB.96.075115} give a slightly lower critical value of $J_{\perp} = 1.12$. At $U=0$, the AFM order parameter goes to zero as $J_\perp\to0$, but it is nonzero at $J_\perp=0$ for any finite value of $U$, as known for the Hubbard model on square lattice. 

Collecting all the critical $J_{\perp}$'s for different 
values of $U$ in Fig.~\ref{fig:AFM}, we obtain the phase diagram shown in Fig.~\ref{fig:phase_diagram_HF}.
The system goes from an antiferromagnet to a Kondo singlet upon increasing $J_\perp$ or $U$.
The Hubbard interaction $U$ favors the Kondo singlet phase, as increasing $U$ at fixed $J_\perp$ brings the system from the AFM to the Kondo singlet phase.  Overall, our phase diagram is in agreement with the one obtained using the  Monte Carlo method~\cite{PhysRevB.70.020402}.
Interestingly, the critical value of $U$ is found to  scale like $1/J_\perp$ when $J_\perp$ is small, i.e., the phase boundary tends towards a finite value of $UJ_\perp$ as $J_\perp\to0$.
This being said, at $J_\perp=0$, the system will always remain an antiferromagnet, as it will always be a Kondo singlet if $J_\perp$ is large enough. The theory of Kondo insulators in Ref.~\cite{PhysRevB.96.075115} provides the following leading equation for the critical boundary between the AFM and Kondo singlet phases: $J_\perp (J_\perp + a U) = b$, where $a$ and $b$ are two positive constants. Indeed, the phase boundary in Fig.~\ref{fig:phase_diagram_HF} looks quite like a parabola.
More generally, $b$ is a function that can be written as a power series in $\frac{1}{(J_\perp+aU)^2}$, while $a$ is a constant. Notably, it also explains why $U$ helps the Kondo singlet. It does so because it adds to $J_\perp$ and acts likewise.
\begin{figure}[htbp]
\centerline{\includegraphics[width=0.99\hsize]{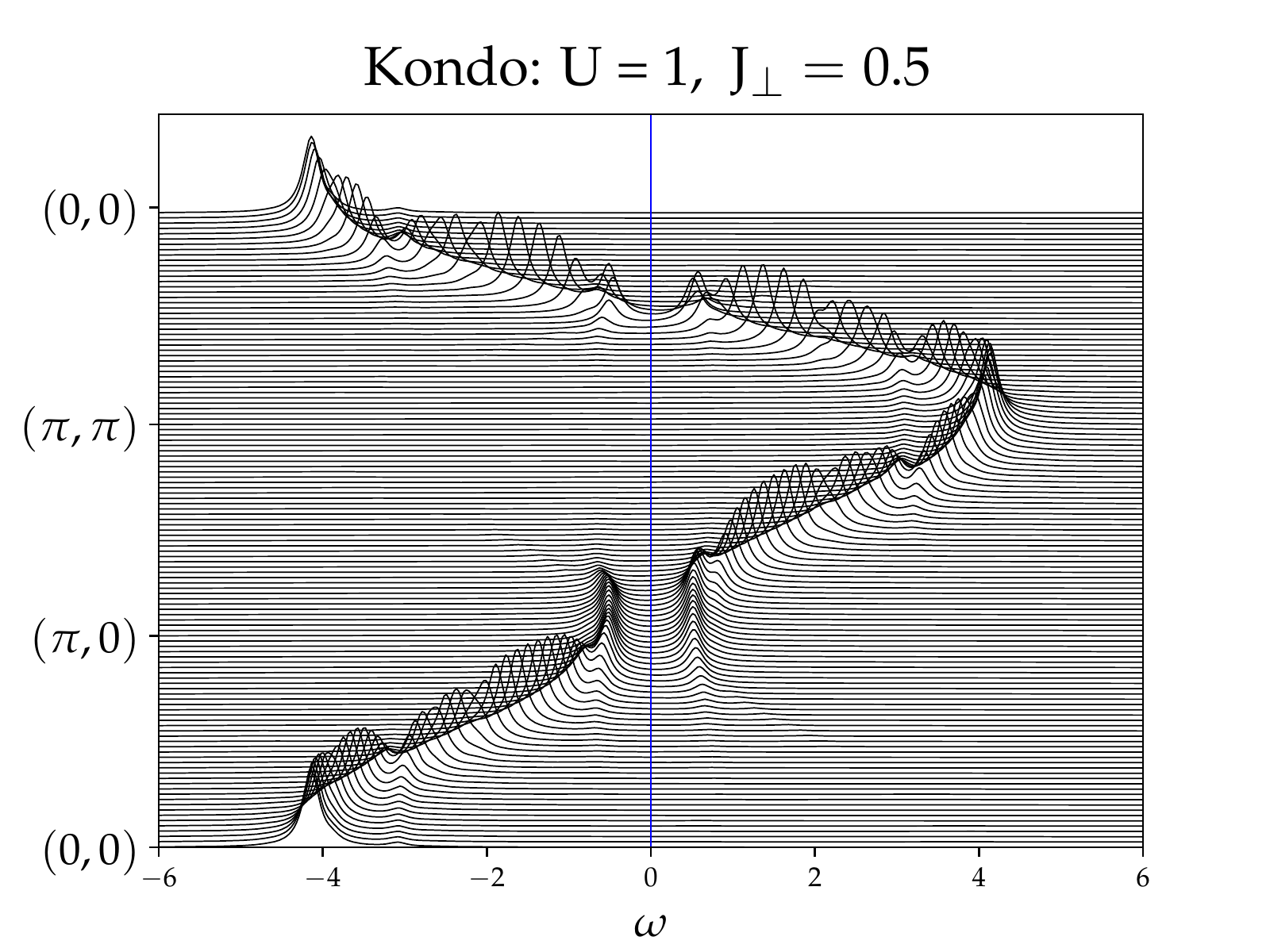}}
\centerline{\includegraphics[width=0.99\hsize]{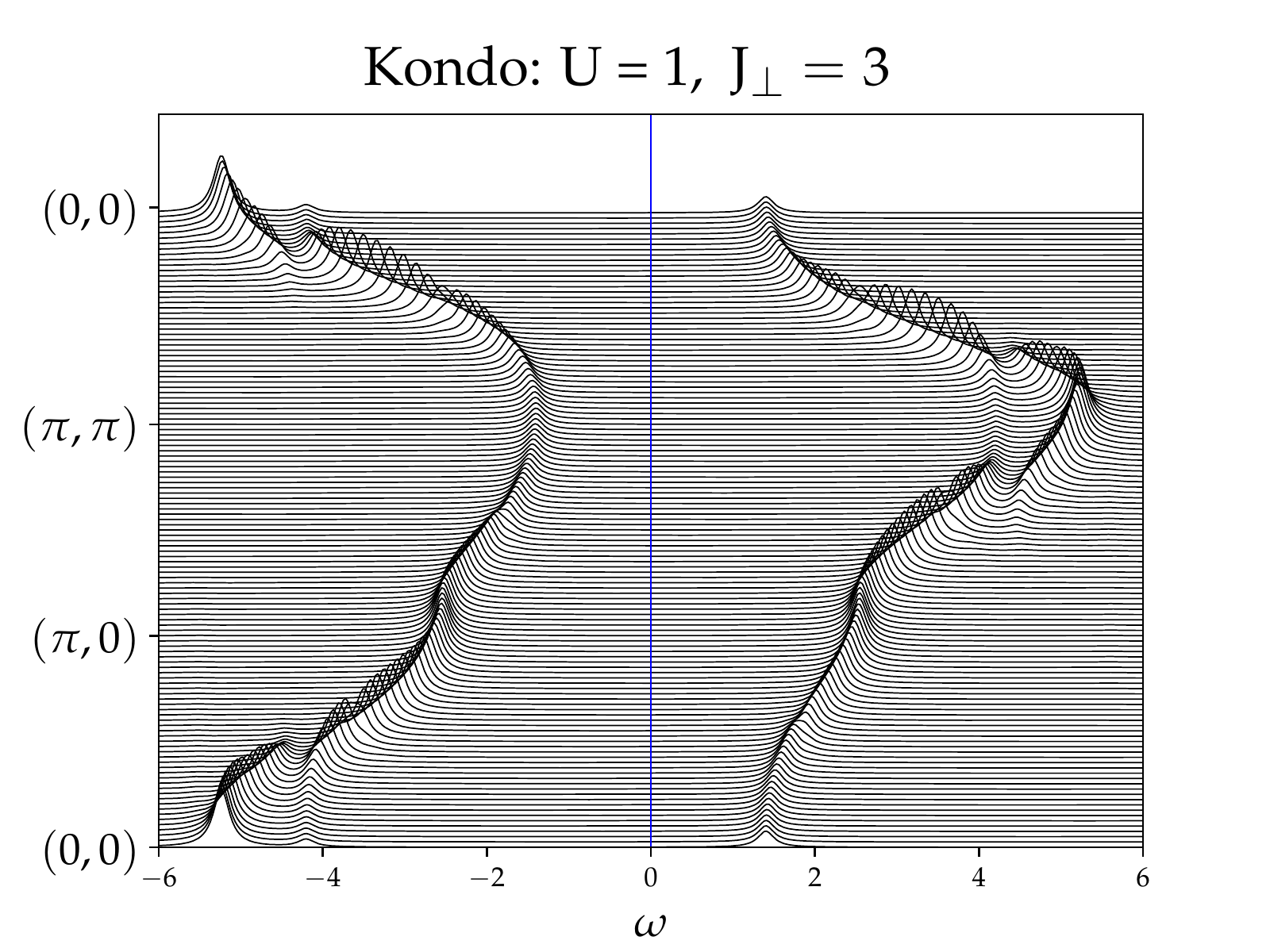}}
\caption{The spectral function in the AFM  (upper panel, $J_\perp = 0.5$) and in the Kondo singlet phase (lower panel, $J_\perp = 3$) at half-filling.
The on-site repulsion $U$ is set to 1 and $t^{\prime} = 0$. The Fermi level is located at zero frequency and the Lorentzian broadening is set to $\eta = 0.12$.\label{fig:spectral}}
\end{figure}

Figure~\ref{fig:gap} shows the charge gap $\Delta_c$ as a function of $1/J_\perp$, for $U=0$, 2, and 4. At a fixed $U$, increasing $J_\perp$ increases the spectral gap. At half-filling, the system is always an insulator for all values of $J_\perp$ and $U$, both in the AFM and the Kondo singlet phases. But depending on the strength of $J_\perp$ (and $U$), the charge gap comes from different points in the Brillouin zone~\cite{PhysRevB.96.075115}.

In  Fig.~\ref{fig:spectral}, we show the spectral function in the two phases at $U = 1$.  In the AFM phase (top panel, $J_\perp = 0.5$), the spectral (one-particle) gap opens along the AFM zone boundary, as expected.
By contrast, in the Kondo singlet phase  (bottom panel, $J_\perp = 3$), the spectral gap is more or less constant across the zone and the spectrum resembles more that of a Mott insulator. 
This is because the charge gap in a Kondo insulator is the cost of destroying a singlet  locally by adding or removing a conduction electron~\footnote{This is different from the spin gap in Kondo insulators, which is the cost of creating a triplet excitation}. Our VCA method does not allow an access to the (two-particle) spin gap, which is expected to vanish in the AFM phase because of Goldstone's theorem, which is exactly like the Mott gap, that is, the cost of adding or removing an electron in the half-filled Hubbard model. Notably, the approach developed by Kumar nicely establishes the similarity between the Kondo and Mott-Hubbard insulators~\cite{PhysRevB.96.075115, PhysRevB.77.205115}. The charge gap in the half-filled Hubbard-Kondo lattice model from our VCA calculations is basically showing the same.

\begin{figure}[htbp]
\centerline{\includegraphics[scale=1]{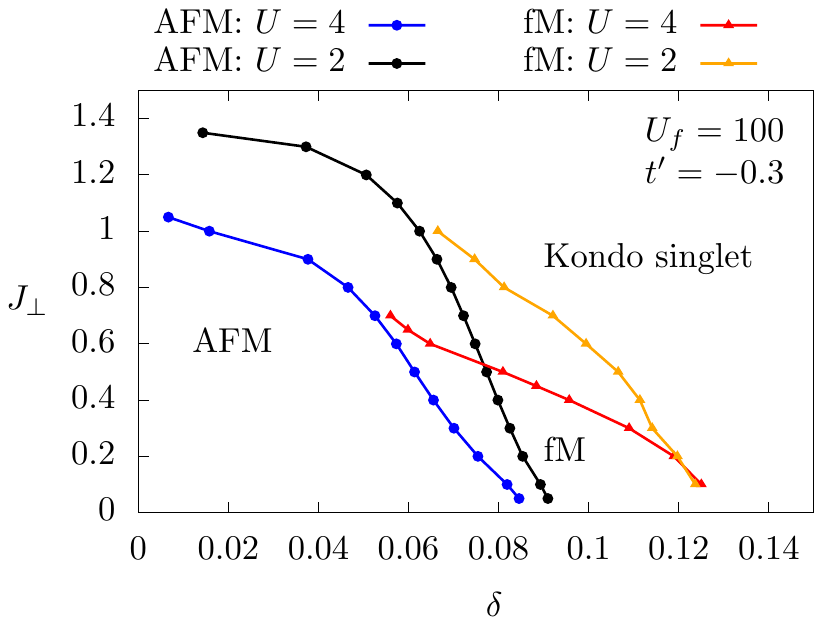}}
\caption{Phase diagram of the Hubbard-Kondo lattice model~\eqref{H_Kondo} in the $(J_{\perp}, \delta)$ plane. 
At moderate $J_{\perp}$, the ground state  is an antiferromagnet at lower doping $\delta$ and a ferromagnet at higher doping, before transiting towards the singlet phase. 
The Kondo singlet phase is stable at large $J_{\perp}$ for all electronic densities. 
The Coulomb interaction is set to $U = 2$ and $U= 4$, and the second-neighbor hopping to $t^{\prime} = -0.3$.\label{fig:AFM-FM}}
\end{figure}
\subsection{Phase diagram at finite doping}
We now push the system away from half-filling, going to small doping $\delta = 1-n$ by changing the number of conduction electrons.
In order to guarantee that the spin susceptibility of the host metallic state at small $\delta$ is peaked at wave vector $ \Qv = (\pi,\pi)$, we add a second-neighbor hopping $t' = -0.3$~\cite{PhysRevB.82.245105}.
Note that the presence of a nonzero $t'$ breaks particle-hole symmetry in the model, which helps bringing the system smoothly away from half-filling in the variational cluster approximation.
The results are presented on Fig.~\ref{fig:AFM-FM} in the $(J_{\perp}, \delta)$ plane.

Close to $\delta=0$, the ground state remains antiferromagnetic.
Upon increasing $\delta$,  the conduction electrons prefer to order ferromagnetically. The localized electrons also do the same. But relative to each other, these two subsystems order antiferromagnetically. However, the net magnetization is non-zero  because the conduction electron mangetization doesn't fully cancels the magnetization of the localized moments. Hence, we like to call this a "ferrimagnetic" (fM) phase. 
Eventually, upon further increasing the exchange interaction, $J_\perp$, the Kondo singlet phase is reached.
The critical doping where the AFM phase disappears completely is about $\delta \approx 0.09$ at  $U = 2$ and $\delta \approx 0.08$ at $U = 4$. 
At finite doping, the transition between the Kondo singlet and magnetically ordered  phases becomes discontinuous (first order).
The  fM phase extends to larger dopings for small $J_\perp$,
since in this limit, 
 the Kondo singlet formation gets weaker and a larger doping does not favour the AFM phase.
It is also clear that introducing the second-neighbor hopping reduces the critical $J_\perp$ towards the Kondo singlet, if we compare the values at $\delta=0$ with the corresponding curves of Fig.~\ref{fig:AFM}.

\section{Conclusion}\label{conclusion}
Using the variational cluster approximation, we have obtained the ground state phase diagram of the Hubbard-Kondo lattice model in the {  $J_\perp$-$U J_\perp$} plane at half-filling, and in the   $\delta$-$J_{\perp}$ plane at finite doping. 
At half-filling,  the model exhibits a continous transition from the N\'eel antiferromagnetic phase for small $J_\perp$'s to the Kondo singlet phase for large $J_\perp$'s with a critical $J_\perp$ that decreases with increasing $U$. The boundary between the two phases is described by the equation, $J_\perp (J_\perp + a U ) = b$, for $a\approx 0.58$ and $b\approx 4.26$.
Away from half-filling, the antiferromagnetic phase survives at small doping, but a  ferrimagnetic phase appears at larger doping and lower $J_\perp$. 
The Kondo singlet phase is stable at strong $J_\perp$. 
The transition from the Kondo singlet to the magnetic phases becomes discontinuous at finite doping.
\begin{acknowledgments}
We gratefully acknowledge conversations with G. Baskaran, B. Svistunov, N. Prokof'ev and M. Boninsegni.  B.K. thanks ICTP for Associate visits, during one of which this work was started, and acknowledges financial support under UPE-II scheme of JNU. Computing resources were provided by Compute Canada and Calcul Qu\'ebec. 
\end{acknowledgments}
%

\end{document}